\documentclass[twocolumn]{aastex631}

\newcommand{\msigma}{$M_{\rm BH}-\sigma_{\star}$}

\newcommand{\mbh}{$M_{\rm BH}$}
\newcommand{\sersic}{S\'{e}rsic}

\newcommand{\msun}{$M_{\odot}$}

\submitjournal{ApJ}

\shorttitle{Reverberation Mapping of IC\,4329A}
\shortauthors{Bentz et al.}

\begin{document}

\title{Reverberation Mapping of IC\,4329A}

\author[0000-0002-2816-5398]{Misty C.\ Bentz}
\affiliation{Department of Physics and Astronomy,
		 Georgia State University,
		 Atlanta, GA 30303, USA}
\email{bentz@astro.gsu.edu}

\author[0000-0003-0017-349X]{Christopher A.\ Onken}
\affiliation{Research School of Astronomy and Astrophysics, 
        Australian National University, 
        Canberra, ACT 2611, Australia}
        
\author[0000-0001-6279-0552]{Rachel Street}
\affiliation{LCOGT, 6740 Cortona Drive, Suite 102, 
        Goleta, CA 93117, USA}

\author[0000-0002-6257-2341]{Monica Valluri}
\affiliation{Department of Astronomy,
         University of Michigan,
         Ann Arbor, MI, 48109, USA}

\begin{abstract}
We present the results of a new reverberation mapping campaign for the
broad-lined active galactic nucleus (AGN) in the edge-on spiral
IC\,4329A.  Monitoring of the optical continuum with $V-$band
photometry and broad emission-line flux variability with
moderate-resolution spectroscopy allowed emission-line light curves to
be measured for H$\beta$, H$\gamma$, and \ion{He}{2} $\lambda 4686$.
We find a time delay of $16.3^{+2.6}_{-2.3}$\,days for H$\beta$, a
similar time delay of $16.0^{+4.8}_{-2.6}$\,days for H$\gamma$, and an
unresolved time delay of $-0.6^{+3.9}_{-3.9}$\,days for
\ion{He}{2}. The time delay for H$\beta$ is consistent with the
predicted value from the relationship between AGN luminosity and broad
line region radius, after correction for the $\sim2.4$\,mag of
intrinsic extinction at 5100\,\AA. Combining the measured time delay
for H$\beta$ with the broad emission line width and an adopted value
of $\langle f \rangle = 4.8$, we find a central supermassive black
hole mass of $M_{\rm BH}=6.8^{+1.2}_{-1.1}\times10^7$\msun.
Velocity-resolved time delays were measured across the broad H$\beta$
emission-line profile and may be consistent with an ``M''-like shape.
Modeling of the full reverberation response of H$\beta$ was able to
provide only modest constraints on some parameters, but does exhibit
agreement with the black hole mass and average time delay.  The models
also suggest that the AGN structure is misaligned by a large amount
from the edge-on galaxy disk.  This is consistent with expectations
from the unified model of AGNs, in which broad emission lines are
expected to be visible only for AGNs that are viewed at relatively
face-on inclinations.

\end{abstract}



\section{Introduction} 

IC\,4329A is a striking example of a broad-lined active galactic
nucleus (AGN) in an edge-on galaxy.  Within the unified model for AGNs
\citep{antonucci93}, this would suggest that the rotation axis of the
central supermassive black hole and the rotation axis of the galaxy
disk are misaligned by a considerable angle.  The misalignment of the
galaxy and black hole are further supported by a $\sim 50^{\circ}$
difference between the position angle of the disk of IC\,4329A and the
position angle of extended radio emission from its nucleus
\citep{schmitt97,nagar99b}.  AGN fueling events can torque the spin
axis of the black hole (e.g., \citealt{king07}) while black hole
mergers can flip the spin axis (e.g., \citealt{lousto16}).  With its
membership within a loose group of galaxies inside the cluster
Abell~3574 \citep{disney73,wilson79,abell89} and close proximity to
the giant lenticular IC\,4329, it is possible that either of these
scenarios may have occurred in IC\,4329A. Furthermore, there is
observational evidence for ongoing interactions between IC\,4329 and
IC\,4329A \citep{read98}, and IC\,4329 is a shell galaxy
\citep{malin83}, which may be evidence of a past merger.

IC\,4329A hosts a highly reddened AGN that is nevertheless apparently bright across the electromagnetic spectrum, which led \citet{wilson79} to suggest that it is the ``nearest quasar''.  Prior studies have concluded that the reddening arises from the AGN structure itself as well as the dust lane of the galaxy disk \citep{wolstencroft95,mehdipour18}.  IC\,4329A thus provides a rare opportunity to carry out detailed studies of  the absorbing material around an accreting supermassive black hole, from the far-infrared through the X-rays (e.g., \citealt{mehdipour18}).  

\begin{figure*}
    \centering
    \plotone{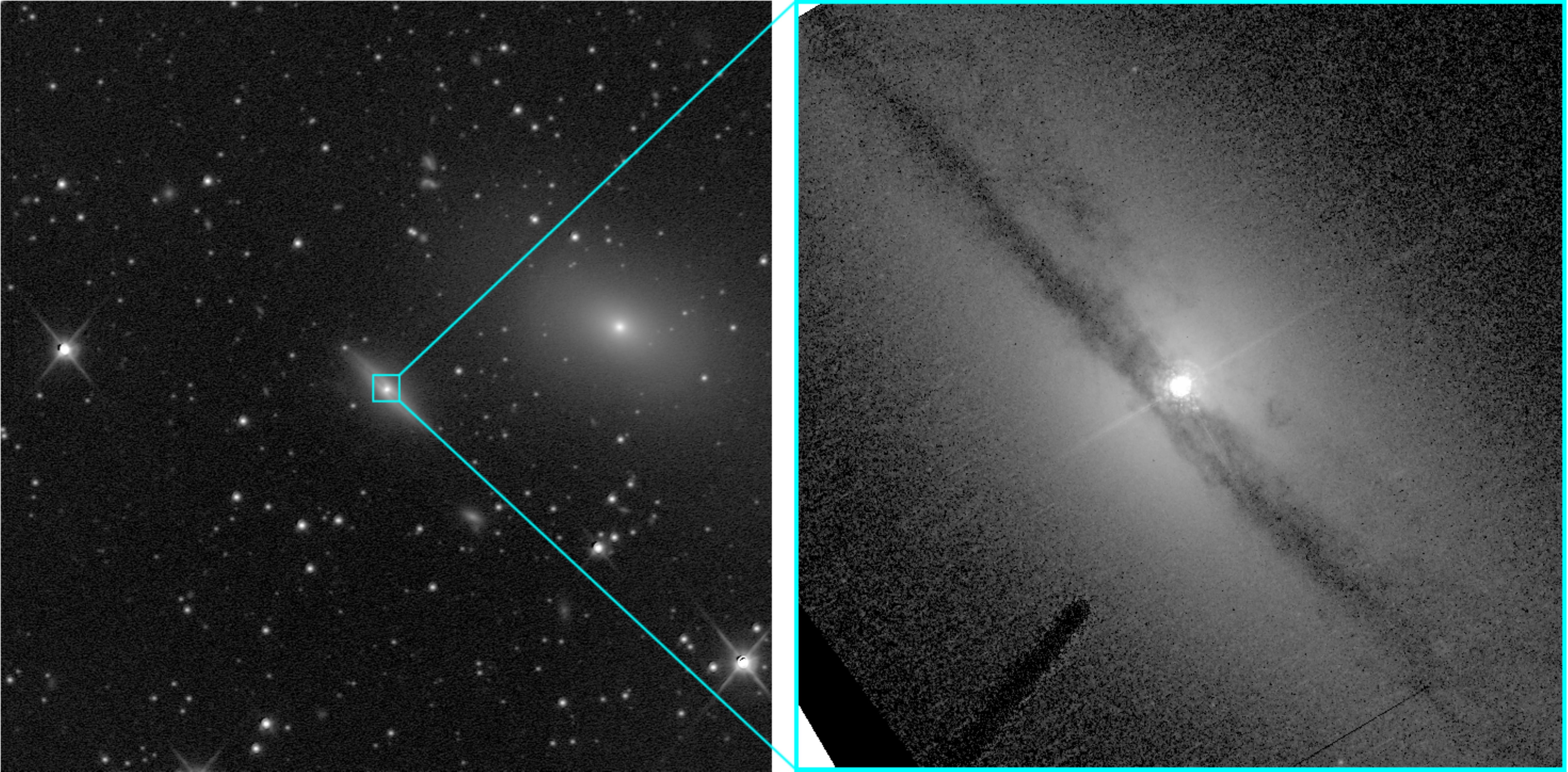}
    \caption{Left: The central $10\arcmin \times 10\arcmin$ of the reference image of IC\,4329A, centered on the AGN. The image is oriented with North up and East to the left, and the giant lenticular galaxy IC\,4329 is visible to the west of IC\,4329A. Right: Hubble Space Telescope image of the central $20\arcsec  \times 20\arcsec$ in IC\,4329A, highlighting the bright central AGN peeking over the dust lane in the disk of the galaxy.}
    \label{fig:refimg}
\end{figure*}

Currently lacking among the studies of IC\,4329A, however, is an accurate constraint on the black hole mass.  Reverberation mapping \citep{cackett21}, which uses spectrophotometric monitoring of an AGN to track changes in the continuum flux and the echoes of those changes in the broad emission lines, is often applied to bright and nearby broad-lined AGNs and has resulted in a sample of $\sim 70$ black hole mass measurements to date in galaxies with $z\lesssim0.4$ \citep{bentz15}.

Strong flux variability in the nucleus of IC\,4329A has been known for decades and an early reverberation mapping attempt was made by \citet{winge96}.  However, that early program was carried out before the typical size of the broad line region (BLR) was well understood, and when estimates were about a factor of 10 too large.  With an average (median) sampling rate of 8 (4) days, and utilizing the CTIO 2D-Frutti camera in the spectrograph — a photon-counting detector with significantly poorer performance than a CCD — the light curves were noisy, undersampled, and provided unreliable time delay measurements \citep{peterson04}.  

We therefore undertook a new reverberation mapping campaign in the first half of 2022 with the aim of constraining the black hole mass of IC\,4329A, and we describe the results of that program in this manuscript. 

\section{Observations}

IC\,4329A is an edge-on spiral galaxy located in the direction of Centaurus at RA=13h49m19.3s, Dec=$-30^{\circ} 18\arcmin 34\arcsec$ (J2000.0), and $z=0.01605$.  Based on fundamental plane studies \citep{hudson01,bernardi02}, the distance to the giant lenticular IC\,4329 is $D = 59\pm9$\,Mpc \citep{tully09}, which is consistent with the distance of $D=69\pm18$\,Mpc derived from observations of a Type 1a supernova  \citep{stahl21}.  IC\,4329A is expected to have a similar distance as IC\,4329 given the signatures of ongoing interactions between them.

\subsection{Imaging}

Photometric monitoring of IC\,4329A was carried out with the Las Cumbres Observatory (LCO; \citealt{brown13}) 1.0\,m telescope network under program NSF-2022A-012 (PI: Bentz)\footnote{LCO telescope time was granted by NOIRLab through the Mid-Scale Innovations Program (MSIP). MSIP is funded by NSF.}.  Observations began on 2022 February 03 (UT) and continued throughout the 2022A semester, ending on 2022 July 31.  A 60\,s $V-$band image was scheduled to occur  every $\sim 8$ hours using the Sinistro imaging cameras, providing a $26\farcm5 \times 26\farcm5$ field of view and an angular scale of 0\farcs389 per pixel.

Over the course of the semester, a total of 374 images were acquired from the LCO facilities at three sites -- 164 from Cerro Tololo Interamerican Observatory (CTIO), 138 from South African Astronomical Observatory (SAAO), and 72 from Siding Spring Observatory (SSO) --
with a median (average) time sampling of 0.35 (0.47) days.  All observations were reduced by the LCO pipeline \citep{mccully18}, which applies typical CCD corrections using bias, flat, and dark frames, and the fully reduced images were downloaded from the LCO archive\footnote{https://archive.lco.global}.

The $V-$band covers a relatively emission-line free region of the spectral energy distribution for nearby AGNs, and is therefore an excellent probe of the continuum variability.  However, nearby AGNs live in optically bright and spatially resolved nearby galaxies.  We therefore relied on image subtraction methods \citep{alard98,alard00} to isolate and measure the variable flux in the nucleus of IC\,4329A.  Image subtraction has the advantage over aperture photometry in that nonvariable flux sources, such as the host galaxy, are removed in the processing.  The removal of the host galaxy results in a reduction of the noise caused by variable seeing affecting the amount of light within a fixed aperture, and it also removes the damping effect on the intrinsic variability that is caused by a large constant-flux offset.  Because of the advantages, image subtraction methods are commonly applied in studies of transients and other time domain phenomena including microlensing \citep{udalski08,udalski15}, supernovae  \citep{riess01,miknaitis07,melinder08}, tidal disruption events \citep{holoien16,brown18}, as well as AGN variability \citep{fausnaugh16,grier17,bentz21a}.  

We first registered all images to a common reference grid with the algorithm of \citet{siverd12}.  Using the algorithms of \citet{alard98} and \citet{alard00}, we then built a reference image (Figure~\ref{fig:refimg}) from the subset of frames collected at CTIO with the lowest backgrounds and best seeing.  For each observation collected throughout the semester, we then convolved the reference frame to match the image, subtracted the convolved reference from the observed image, and measured the residual counts at the position of the AGN using an aperture with a radius of 9 pixels.

To convert the residual counts to calibrated fluxes, we carried out a
2D surface brightness decomposition of the reference frame.  Using
{\tt Galfit} \citep{peng02,peng10}, we modeled the separate
photometric components of IC\,4329A and the nearby lenticular
IC\,4329.  A model of the point spread function (PSF) of the image was
generated by fitting four Gaussian profiles to a bright and isolated
field star.  For each of the Gaussian profiles, we allowed the first
mode of the Fourier perturbation series to vary in an effort to better
account for slight asymmetries in the PSF shape.  The resultant PSF
model image was then used to model the AGN as well as several field
stars with catalogued $V-$band magnitudes.  An exponential profile
with a radial truncation was used to model the edge-on galaxy disk and
its extinguishing dust lane, while \sersic\ profiles modeled the bulge
and companion galaxy. Finally, the background sky was modeled with a
tilted plane to allow for gradients in both the {\tt x} and {\tt y}
directions.  The final zeropoint of the flux calibration was set by
minimizing the difference between the fitted magnitudes of the fields
stars and their reported magnitudes in the AAVSO Photometric All Sky
Survey catalog \citep{henden14}.  Once the models were optimized, the
final modeled AGN magnitude was combined with the residual counts to
determine the calibrated fluxes for each $V-$band measurement
throughout the campaign.

As we found in our previous monitoring study of NGC\,3783 \citep{bentz21a}, the images acquired from the three Southern LCO 1\,m sites have slight photometric offsets between them.  We identified observations that were collected close in time ($<0.5$\,day) from CTIO and SSO, as well as from CTIO and SAAO.  We then determined and applied the slight linear scaling needed to bring the SAAO and SSO measurements into agreement with the CTIO measurements, as the CTIO data set was used to build the reference frame for image subtraction and to determine the AGN mean magnitude.

\begin{figure}
    \centering
    \epsscale{1.15}
    \plotone{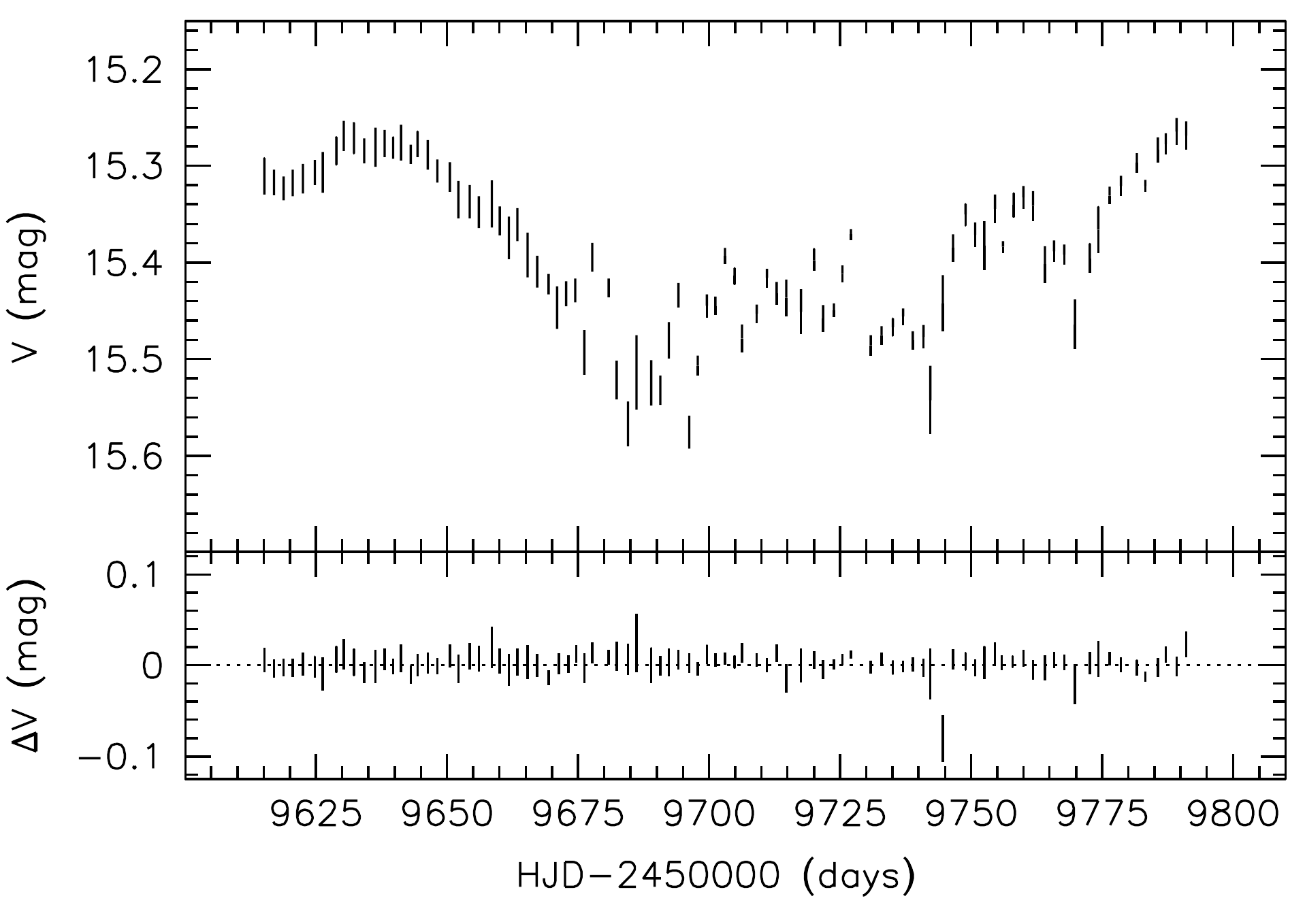}
    \caption{Top: Final calibrated and binned $V-$band light curve of IC\,4329A.  Bottom: Magnitude residuals for a non-varying $V=15.3$\,mag field star. }
    \label{fig:contlc}
\end{figure}

Previous work has found that the flux uncertainties associated with image subtraction methods are often underestimated \citep{zebrun01,hartman05}.  We therefore followed the basic procedure outlined by \citet{hartman04} and examined the residual counts left behind by non-varying field stars, identifying the factor by which the uncertainties needed to be expanded to account for the scatter in residual flux from stars of a similar magnitude as the AGN.  The uncertainties on the AGN fluxes were then inflated by this scale factor.  

Finally, over the course of $\sim 2$ months in the middle of the campaign, the fluxes measured for the AGN varied strongly and extremely rapidly from observation to observation.  Inspection of the images acquired during these dates found that some of this rapid variability coincided with bright and uneven backgrounds in the images that could not be approximated with a simple gradient.  Many of these dates also coincide with the Full Moon, and the uneven backgrounds appear to be associated with the reflection of moonlight within the domes.  Rather than discard all of these frames and accept gaps of several days in the middle of the light curve, we instead discarded only the most egregious frames and binned all the remaining measurements with a bin size of 1.5\,days.  This bin size was determined through testing to provide a reasonable compromise between retaining temporal sampling and decreasing wild flux variations that cause difficulties in the time delay analysis. 

In Figure~\ref{fig:contlc} we display the final calibrated and binned $V-$band light curve of IC\,4329A, along with a plot of the residuals from a field star of similar brightness for comparison.

\subsection{Spectroscopy}

\begin{figure}
    \centering
    \epsscale{1.15}
    \plotone{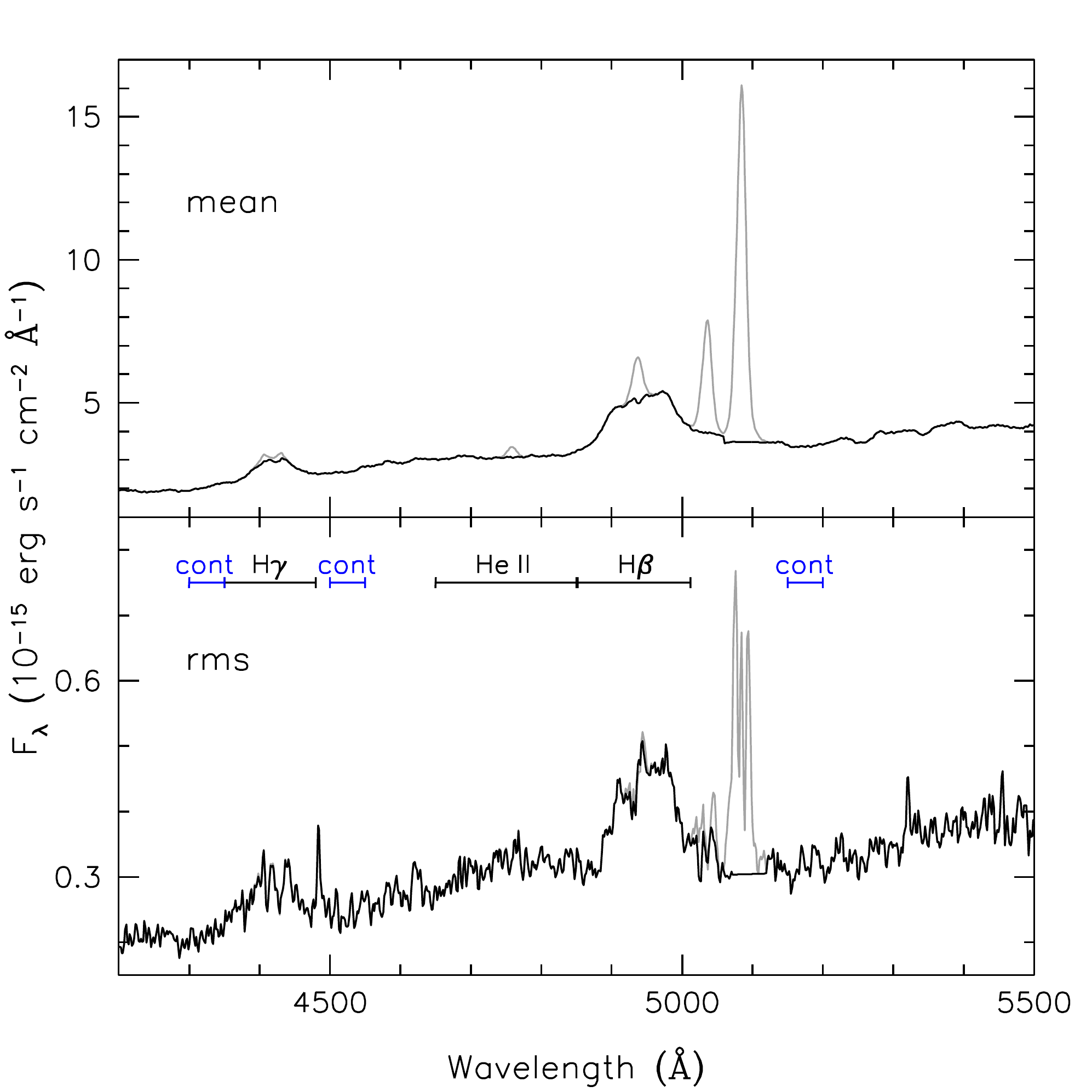}
    \caption{Mean (top) and rms (bottom) of the IC\,4329A spectra collected throughout the monitoring program.  Integration limits for the broad emission lines are displayed as the horizontal bars,  and the adopted continuum windows are displayed as the blue horizontal bars. The gray (black) lines show the spectra before (after) narrow-line subtraction.}
    \label{fig:meanspec}
\end{figure}

Spectroscopic monitoring with the FLOYDS spectrographs on the 2.0\,m Faulkes Telescopes North and South was carried out under programs NSF-2022A-012 and ANU-2022A-002.  The FLOYDS spectrographs are cross-dispersed instruments providing wavelength coverage of $540-1000$\,nm with a dispersion of 3.51\,\AA/pix in the first order, and wavelength coverage of $320-570$\,nm  with a dispersion of 1.74\,\AA/pix in the second order. Observations began on 2022 March 07 and were scheduled every $\sim 24$\,hours  through 2022 July 31.  Over the course of the semester, 45 spectra (23 from FTN and 22 from FTS) were acquired with a median (average) time sampling of 2.0 (3.2) days.  

Each observation consisted of a 1200\,s spectrum of IC\,4329A through the 6.0\arcsec\ slit, with an arc lamp and flat field taken at the same position.  On several visits we also acquired a 60\,s spectrum of the nearby A0 spectrophotometric standard star CD-32 9927 \citep{hamuy92,hamuy94} through the 6.0\arcsec\ slit.

The LCO pipeline splits the two spectral orders into separate images, after which it rectifies each order and  applies typical CCD reductions including biases, flats, and darks as well as an initial wavelength and flux calibration based on cataloged wavelength and sensitivity functions.  We began our custom processing with the 2D flux- and wavelength-calibrated frames produced by the pipeline, focusing on the 2nd order (blue) spectra with their higher spectral resolution and coverage of the H$\beta$ and [\ion{O}{3}] emission. Working in {\tt IRAF}\footnote{IRAF is distributed by the National Optical Astronomy Observatories, which are operated by the Association of Universities for Research in Astronomy, Inc., under cooperative agreement with the National Science Foundation.} we detected and cleaned the science frames of cosmic rays and then extracted 1D spectra of IC\,4329A, the standard star, and the arc lamps with a 10\,pixel extraction width (corresponding to an angular width of 3\farcs37).  We then used the arc lamp spectra to improve the wavelength calibration for each observation, and finally adopted and applied a common dispersion of 1.6\,\AA/pix.

For the visits that included an observation of the standard star, we fit a low order polynomial to the standard star spectrum and compared this to the tabulated spectrum of \citet{hamuy92,hamuy94}.  We then determined the scaling needed to match the tabulated standard spectrum, and applied this scaling to the spectrum of IC\,4329A that was collected in the same visit.  Given the lack of standard star observations on several nights and the nonphotometric conditions under which many spectra were acquired, we were unable to improve the flux calibration of all the observations of IC\,4329A in this way.  We instead focused on determining the absolute calibration of the [\ion{O}{3}] $\lambda 5007$\,\AA\ emission line from photometric nights that included observations of both IC\,4329A and CD-32 9927, finding $F(5007)=(2.34\pm0.13)\times 10^{-13}$\,erg\,s$^{-1}$\,cm$^{-2}$.  This value agrees well with the flux reported by \citet{morris88} of $F(5007)=2.4\times 10^{-13}$\,erg\,s$^{-1}$\,cm$^{-2}$ and is slightly higher, but within 2$\sigma$, of the value reported by \citet{winge96} of $F(5007)=(1.87\pm0.22)\times 10^{-13}$\,erg\,s$^{-1}$\,cm$^{-2}$.

Finally, we applied the spectral scaling method of \citet{vangroningen92} to all of the individual spectra of IC\,4329A, focusing on the [\ion{O}{3}] doublet.  The [\ion{O}{3}] emission lines are not variable on the timescales probed within a single observing semester \citep{peterson13} and may therefore be used as an internal calibration ``lamp''.  The algorithm applies small shifts and smoothing to minimize the differences between an observed spectrum and a reference spectrum (created by averaging together several of the best-quality spectra), thus correcting for residual wavelength shifts, offsets in the flux calibration relative to the value of $F(5007)$ that we determined above, and differences in the spectral resolution arising from variable seeing throughout the campaign. 

In Figure~\ref{fig:meanspec} we display the mean and root-mean-square (rms) of the blue spectra of IC\,4329A collected throughout the monitoring campaign.  Strong variability is evident in the rms of the broad Balmer lines H$\beta$ and H$\gamma$ as well as the \ion{He}{2} $\lambda 4686$\,\AA\ line.

\begin{deluxetable*}{lcccccccccc}
\tablecolumns{11}
\tablewidth{0pt}
\tablecaption{Continuum and Emission-Line Light Curves}
\tablehead{
\multicolumn{3}{c}{Continuum} &
\colhead{} &
\multicolumn{7}{c}{Emission Lines} \\
\cline{1-3}
\cline{5-11}
\colhead{} &
\multicolumn{2}{c}{$V$} &
\colhead{} &
\colhead{} &
\multicolumn{2}{c}{H$\beta$} &
\multicolumn{2}{c}{H$\gamma$} &
\multicolumn{2}{c}{\ion{He}{2}} \\
\colhead{HJD} &
\colhead{$F_{\lambda}$} &
\colhead{$\sigma_F$} &
\colhead{} &
\colhead{HJD} &
\colhead{$F$} &
\colhead{$\sigma_F$} &
\colhead{$F$} &
\colhead{$\sigma_F$} &
\colhead{$F$} &
\colhead{$\sigma_F$} 
}
\startdata
9615.1116 & 2.666 & 0.046 && 9646.0925 & 3.124 & 0.031 & 0.700 & 0.025 & 0.746 & 0.031 \\
9616.9787 & 2.651 & 0.032 && 9647.0745 & 3.075 & 0.029 & 0.680 & 0.023 & 0.747 & 0.029 \\
9618.7941 & 2.636 & 0.031 && 9649.1018 & 3.020 & 0.033 & 0.639 & 0.027 & 0.584 & 0.034 \\
9620.5635 & 2.650 & 0.034 && 9651.0854 & 3.155 & 0.028 & 0.711 & 0.022 & 0.715 & 0.028 \\
9622.5523 & 2.660 & 0.038 && 9657.0617 & 3.022 & 0.032 & 0.685 & 0.030 & 0.521 & 0.036 
\label{tab:lc}
\enddata 

\tablecomments{Heliocentric Julian dates are provided as HJD$-2450000$ (days).  $V-$band flux densities have units of $10^{-15}$\,erg\,s$^{-1}$\,cm$^{-2}$\,\AA$^{-1}$ while emission-line fluxes have units of $10^{-13}$\,erg\,s$^{-1}$\,cm$^{-2}$.  Table~\ref{tab:lc} is published in its entirety in the machine-readable format. A portion is shown here for guidance regarding its form and content.}
\end{deluxetable*}

\begin{deluxetable*}{lccccccc}
\tablecolumns{8}
\tablewidth{0pt}
\tablecaption{Light Curve Statistics}
\tablehead{
\colhead{Time Series} &
\colhead{$N$} &
\colhead{$\langle \Delta T \rangle$ (days)} &
\colhead{$\Delta T_{\rm med}$ (days)} &
\colhead{$\langle F \rangle$} &
\colhead{$\langle \sigma_F / F \rangle$} &
\colhead{$F_{\rm var}$} &
\colhead{$R_{\rm max}$}\\
\colhead{(1)} &
\colhead{(2)} &
\colhead{(3)} &
\colhead{(4)} &
\colhead{(5)} &
\colhead{(6)} &
\colhead{(7)} &
\colhead{(8)} 
}
\startdata
V           & 91 & $2.0 \pm 0.4$ & 1.9 & $2.49 \pm 0.18$  & 0.014  & 0.07 & $1.332 \pm 0.027$ \\
H$\beta$    & 45 & $3.2 \pm 3.9$ & 2.0 & $2.57 \pm 0.25$  & 0.011  & 0.10 & $1.436 \pm 0.028$ \\
H$\gamma$   & 45 & $3.2 \pm 3.9$ & 2.0 & $0.52 \pm 0.08$  & 0.049  & 0.15 & $1.859 \pm 0.126$ \\
\ion{He}{2} & 45 & $3.2 \pm 3.9$ & 2.0 & $0.44 \pm 0.13$  & 0.079  & 0.28 & $6.553 \pm 2.661$ 
\label{tab:lcstats}
\enddata 

\tablecomments{$V-$band flux densities have units of $10^{-15}$\,erg\,s$^{-1}$\,cm$^{-2}$\,\AA$^{-1}$ while emission-line fluxes have units of $10^{-13}$\,erg\,s$^{-1}$\,cm$^{-2}$.}
\end{deluxetable*}

\section{Analysis} \label{sec:analysis}

\subsection{Emission-Line Light Curves and Time Delays} \label{sec:delays}

With our final scaled spectra, we were able to measure the integrated fluxes of the broad emission lines and create their light curves.  We began by setting a local linear continuum based on line-free regions on either side of a broad emission line of interest, and then integrating all of the flux above the continuum within a specified window.  Thus the emission-line fluxes include the narrow line components as constant flux offsets.  While simplistic, this method avoids inserting additional uncertainties into the measurements and has been shown to work well when applied to high signal-to-noise spectra with relatively unblended emission lines (e.g., \citealt{peterson04,bentz09c}).

\begin{figure}
    \centering
    \epsscale{1.15}
    \plotone{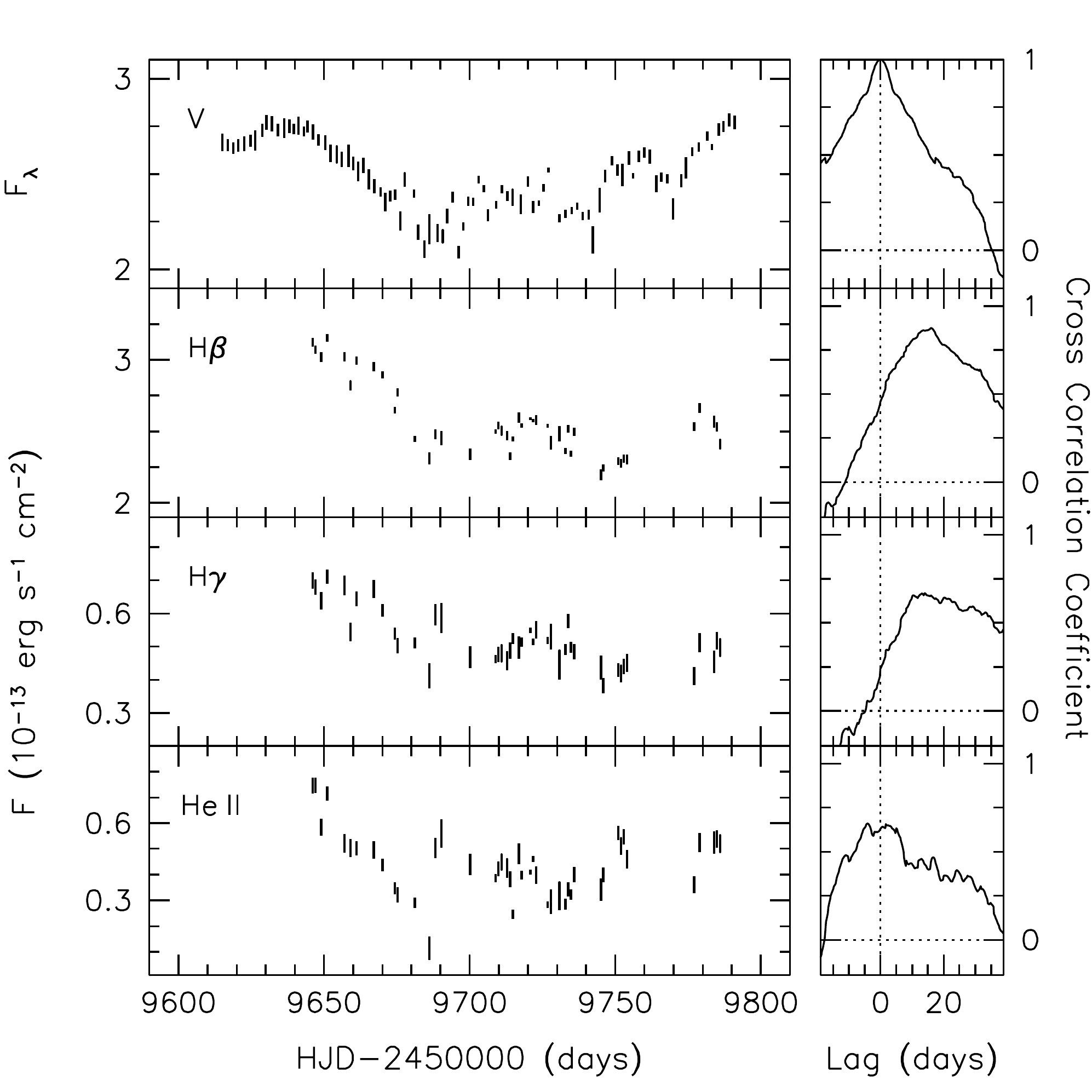}
    \caption{Left: $V-$band and emission-line light curves for
      IC\,4329A. Flux densities are in units of 
        $10^{-15}$\,erg\,s$^{-1}$\,cm$^{-2}$\,\AA$^{-1}$, and
      integrated fluxes are in units of
      $10^{-13}$\,erg\,s$^{-1}$\,cm$^{-2}$.  Right: Cross-correlation
      functions relative to the $V-$band light curve (for the $V-$band
      light curve, this is the autocorrelation function).}
    \label{fig:lc_ccf}
\end{figure}

We provide the light curves in Table~\ref{tab:lc} and plot them in
Figure~\ref{fig:lc_ccf}, while we tabulate several useful statistics
for each light curve in Table~\ref{tab:lcstats}.  For each spectral
feature listed in column (1), we give the number of measurements in
column (2), and the average and median temporal sampling,
respectively, in columns (3) and (4). Column (5) lists the mean flux
and standard deviation, while column (6) lists the mean fractional
error. Column (7) gives the  noise-corrected fractional
  variation, which is computed as
\begin{equation}
    F_{\rm var} = \frac{\sqrt{\sigma^2 - \delta^2}}{\langle F \rangle}
\end{equation}
\noindent where $\sigma^2$ is the variance of the fluxes, $\delta^2$ is their mean-square uncertainty, and $\langle F \rangle$ is the mean flux. Finally, column (8) lists the ratio of
the maximum to the minimum flux in the light curve.

To measure the time delays between the emission-line light curves and the $V-$band light curve, we began by employing the Interpolated Cross Correlation Function (ICCF) method of \citet{gaskell86,gaskell87} with the modifications of \citet{white94}.  The ICCF method determines the cross-correlation function twice, first when the continuum light curve is interpolated and then when the emission-line light curve is interpolated, and then averages the two together.  These averaged CCFs are displayed in the panels on the right side of Figure~\ref{fig:lc_ccf}.

From the CCF, the time delay may be reported as $\tau_{\rm peak}$, the time lag at which the peak ($r_{\rm max}$) of the CCF occurs, or $\tau_{\rm cent}$, the centroid of the CCF  above some value (usually $0.8r_{\rm max}$).  We employ the flux randomization/random subset sampling (FR/RSS) method \citep{peterson98b,peterson04} to quantify the uncertainties on the time delays.  In short, the flux randomization accounts for the effects of the measurement uncertainties on the derived time delay, and the random subset sampling accounts for the inclusion or exclusion of any specific data point in the analysis. The FR/RSS method is run in a Monte Carlo fashion, carrying out a large number of realizations to build up a distribution of CCF measurements.  We ran 1000 realizations of the FR/RSS method and adopted the median and 68\% confidence interval of the distributions to define the time delays, $\tau_{\rm cent}$ and $\tau_{\rm peak}$, and their uncertainties, which are tabulated in Table~\ref{tab:lagwidth}.

\begin{deluxetable*}{lccccccc}
\renewcommand{\arraystretch}{1.2}
\tablecolumns{8}
\tablewidth{0pt}
\tablecaption{Emission-Line Time Lags and Widths}
\tablehead{
\colhead{} &
\colhead{} &
\colhead{} &
\colhead{} &
\multicolumn{2}{c}{mean} &
\multicolumn{2}{c}{rms}\\
\colhead{Line} &
\colhead{$\tau_{\rm cent}$} &
\colhead{$\tau_{\rm peak}$} &
\colhead{$\tau_{\rm jav}$} &
\colhead{FWHM} &
\colhead{$\sigma_{\rm line}$} &
\colhead{FWHM} &
\colhead{$\sigma_{\rm line}$}\\
\colhead{} &
\colhead{(days)} &
\colhead{(days)} &
\colhead{(days)} &
\colhead{(km s$^{-1}$)} &
\colhead{(km s$^{-1}$)} &
\colhead{(km s$^{-1}$)} &
\colhead{(km s$^{-1}$)}
}
\startdata
H$\beta$    & $	16.33^{+2.59}_{-2.28}$ & $	16.00^{+0.50}_{-2.75}$  & $16.09^{+0.54}_{-1.26}$ & $6944\pm51$  & $2247\pm8$  &  $4789\pm869$   & $2112\pm93$   \\
H$\gamma$   & $	16.04^{+4.83}_{-2.60}$ & $	13.25^{+4.25}_{-2.75}$ & $15.48^{+0.96}_{-2.23}$ & $4176\pm162$ & $1703\pm61$ &  $4607\pm948$  & $1908\pm149$   \\
\ion{He}{2} & $	-0.59^{+3.87}_{-3.85}$ & $	0.25^{+4.75}_{-4.75}$   & $-1.37^{+2.04}_{-2.37}$ & \nodata      & \nodata     &  $8099\pm1574$  & $3334\pm134$ 
\label{tab:lagwidth}
\enddata 

\tablecomments{Reported time lags are in the observer's frame, while line widths are  in the rest frame.}
\end{deluxetable*}

We also investigated the time delays with {\tt JAVELIN} \citep{zu11}, which fits a damped random walk model to the continuum light curve and then determines the best parameters for a top-hat function with which to delay and smooth the continuum light curve to match the emission-line light curve. {\tt JAVELIN} is capable of fitting multiple emission lines at the same time, and it assesses the uncertainties on the reported time delay through a Bayesian Markov Chain Monte Carlo method.  We fit all of the emission lines simultaneously, as well as individually, and found that both methods recovered time delays that agree fairly well.  We list the time delays, denoted as $\tau_{\rm jav}$, from fitting all three emission lines simultaneously in Table~\ref{tab:lagwidth}.

\subsection{Line Width Measurements}

The widths and shapes of the broad emission lines provide information
on the kinematics of the line-emitting gas, and are generally reported
as the velocity dispersion of the line profile, $\sigma_{\rm
  line}$, and the full-width at half maximum, FWHM.  Previous
reverberation studies have often reported different broad-line
widths when measured in the mean vs.\ the rms spectrum (for a
compilation, see \citealt{bentz15}), with the rms spectrum crucially
providing a measure of the kinematics of the reverberating gas.
Furthermore, the rms spectrum suppresses the non-variable 
  narrow-line contribution.

While the rms spectrum suppresses the narrow line contribution, the narrow lines must be fitted and removed from the mean spectrum before measuring the broad line widths.  We used the [\ion{O}{3}] $\lambda 5007$ line as a template that was shifted and scaled to match the other narrow lines in the relevant portions of the spectrum  --- [\ion{O}{3}] $\lambda 4959$ and $\lambda 4363$, and the narrow components of H$\beta$, \ion{He}{2}, and H$\gamma$ --- and then subtracted.  In Figure~\ref{fig:meanspec} we display the mean and rms spectra before and after narrow-line subtraction in gray and black, respectively.  As expected, the rms profiles of the broad emission lines are relatively unaffected by the narrow-line subtraction.

To determine the uncertainties on the line widths, we employed a bootstrap sampling method that selects a random subset of spectra, creates a mean and rms spectrum, and then measures the line width directly from the data by fitting a local linear continuum under each emission line and then constraining the width of the flux in excess of the continuum.  We ran 1000 realizations of the method and built up a distribution of $\sigma_{\rm line}$ and FWHM measurements, from which we report the median and 68\% confidence interval as the measurement and its uncertainty, respectively. As expected, the line widths in the mean spectrum increased after narrow-line subtraction had been applied (especially FWHM), while the line widths remained the same in the rms spectrum before and after narrow-line subtraction.  We report both FWHM and $\sigma_{\rm line}$ for the mean and rms spectra, after narrow-line subtraction, in Table~\ref{tab:lagwidth}. 

All reported line widths were corrected for the resolution of the spectrograph by assuming
\begin{equation}
    \Delta \lambda_{\rm obs}^2 \approx \Delta \lambda_{\rm true}^2 + \Delta \lambda_{\rm disp}^2
\end{equation}
\noindent where $\Delta \lambda_{\rm obs}$ is the measured width of an emission line and $\Delta \lambda_{\rm disp}$ is the broadening caused by the instrument.  We estimated $\Delta \lambda_{\rm true}$, the intrinsic line width, by adopting the width of [\ion{O}{3}] $\lambda 5007$\,\AA\ measured by \citet{whittle92} through a small spectrograph slit and at high resolution, FWHM$=550$\,km\,s$^{-1}$. Combining $\Delta \lambda_{\rm true}$  with $\Delta \lambda_{\rm obs}$, we found $\Delta \lambda_{\rm disp}=13.9$\,\AA.

\section{Discussion}

\subsection{Radius vs.\ Luminosity}

The mean and rms flux at $5100\times(1+z)$\,\AA\ throughout the monitoring campaign was measured to be $F_{\lambda}(5100)=(3.49\pm0.30)\times10^{-15}$\,erg\,s$^{-1}$\,cm$^{-2}$\,\AA$^{-1}$.  However, with the large spectroscopic aperture employed, a significant fraction of this flux is expected to come from stars in the host galaxy rather than the AGN itself.  IC\,4329A was observed with the Hubble Space Telescope Advanced Camera for Surveys High Resolution Channel through the F550M filter in 2006 as part of program GO-10516 (see Figure~\ref{fig:refimg}).  \citet{bentz09b} report a 2D surface brightness decomposition of the image, from which we determined the host galaxy contribution through the spectroscopic aperture utilized in this work, $F_{\lambda,\rm gal}(5100)= (2.15\pm0.22) \times 10^{-15}$\,erg\,s$^{-1}$\,cm$^{-2}$\,\AA$^{-1}$.

Removing the starlight contribution to the flux measured at $5100\times(1+z)$\,\AA, correcting for Galactic absorption ($A_B=0.21$\,mag, \citealt{schlafly11}), and assuming a distance of $D=59\pm9$\,Mpc (as measured for IC\,4329, \citealt{tully09}), we find an average nuclear luminosity of $\log \lambda L_{\lambda}(5100) = 42.53\pm0.15$\,L$_{\odot}$ for IC\,4329A during this monitoring campaign.  Based on the relationship between BLR radius and AGN luminosity for local Seyferts \citep{bentz13}, we would thus predict an H$\beta$ time delay of $\sim 6$\,days.

However, the observed nuclear luminosity is likely a severe underestimate because of the intrinsic extinction along our line of sight to the center of IC\,4329A. \citet{mehdipour18} analyzed the spectral energy distribution of IC\,4329A from the X-rays through the far-infrared, finding that the internal extinction was best described by a flat or gray extinction curve, similar to that described by \citet{czerny04}, with $E(B-V)=1.0\pm0.1$\,mag. These results suggest that the intrinsic AGN flux at $5100\times(1+z)$\,\AA\ is $\sim 9.5\times$ larger than the observed flux.  Adjusting the nuclear luminosity by this factor then predicts an H$\beta$ time delay of $\sim 18$\,days, in good agreement with the measurements we present here given the uncertainties involved in correcting internal extinction.

\subsection{Black Hole Mass} \label{sec:mbh}

The time delay for a broad emission line ($\tau$) gives a measure of the light-crossing time and thus the physical size of the emission-line region, while the width of the line ($V$) provides a constraint on the line-of-sight velocity of the gas in the region.  The two may be combined to derive the black hole mass as:
\begin{equation}
    M_{\rm BH} = f \frac{c\tau V^2}{G}
\end{equation}
\noindent where $c$ is the speed of light, and $G$ is the gravitational constant.  The scale factor $f$ is an order unity term that includes the detailed geometry and kinematics of the BLR as well as the inclination angle at which the AGN system is viewed.

Historically, it has been difficult to constrain the BLR structure and
kinematics for most AGNs that have been studied with reverberation
mapping.  Thus, a population-average scale factor, $\langle f
\rangle$, is generally adopted to bring the sample of
reverberation-based masses into general agreement with samples of
black hole masses from dynamical modeling through comparison of their
\msigma\ relationships.  Previous studies have found values of
$\langle f \rangle=2.8-5.5$, depending on the details of the adopted
samples and the fitting method
\citep{onken04,graham11,park12,grier13,batiste17b}.  These values of
$\langle f \rangle$ are appropriate for time delays given by
$\tau_{\rm cent}$ and line widths given by $\sigma_{\rm line}$(rms),
the combination of measurements that has been shown to provide the
smallest scatter among black hole masses derived from multiple
reverberation experiments targeting the same AGN \citep{peterson04}.
We note, however, that while adopting $\langle f \rangle$ minimizes
any bias in a sample of reverberation masses, it also means that the
reverberation mass for any single AGN may be uncertain by a factor of
a few.  In particular, the inclinations of individual AGNs to our line
of sight are expected to be random, and the analysis of
\citet{williams18} suggests that inclination plays a significant role
in the variance of individual $f$ values determined for different AGNs
(see also, e.g., \citealt{collin06}).

Here we adopt $\langle f \rangle=4.8$ from \citet{batiste17b} because
of their careful assessment of galaxy morphology on the determination
of $\sigma_{\star}$ among the AGN sample.  Combined with
  $\tau_{\rm cent}$ and $\sigma_{\rm line}$(rms) for H$\beta$, we
find $M_{\rm BH}=6.8^{+1.2}_{-1.1}\times10^7$\msun.   We find a
  similar mass with slightly larger uncertainties, $M_{\rm
    BH}=5.5^{+1.9}_{-1.2}\times10^7$\msun, using the measurements of
  H$\gamma$ and adopting the same value of $\langle f \rangle$.

The \msigma\ relationship is commonly used to estimate black hole
masses in galaxies, although estimates for the black hole mass in
IC\,4329A based on the \msigma\ relationship are likely to be biased
somewhat high because of the rotational contribution of the edge-on
galaxy disk to $\sigma_{\star}$.  Using the \msigma\ relationship of
\citet{tremaine02}, \citet{markowitz09} estimated $M_{\rm
  BH}=2^{+2}_{-1}\times10^8$\msun\ for IC\,4329A, which is indeed
somewhat higher than the reverberation mass we present
here. \citet{markowitz09} also estimated the black hole mass using the
\citet{mchardy06} relationship between bolometric luminosity, black
hole mass, and X-ray power spectral density break, finding $M_{\rm
  BH}=1.3^{+1.0}_{-0.3}\times10^8$\msun.  \citet{markowitz09} assumed
a luminosity distance of 78.6\,Mpc when calculating the bolometric
luminosity, so adopting the measured distance to IC\,4329 of
$D=59$\,Mpc decreases the predicted mass by $\sim23$\%, bringing it
into better agreement with the reverberation mass.  Finally,
\cite{ponti12} explored the relationship between X-ray excess variance
and \mbh, using the reverberation sample to define a scaling
relationship between the two. With this scaling relationship, they
predict $M_{\rm BH}=2.1^{+1.3}_{-0.8} \times 10^8$\msun\ for
IC\,4329A, although we note that they adopted a mass estimate for
IC\,4329A based on \msigma\ and included it as one of the few black
holes with $M_{\rm BH}>10^8$\msun\ in the reverberation sample that
defined the scaling relationship.

\begin{figure}
    \centering
    \epsscale{1.15}
    \plotone{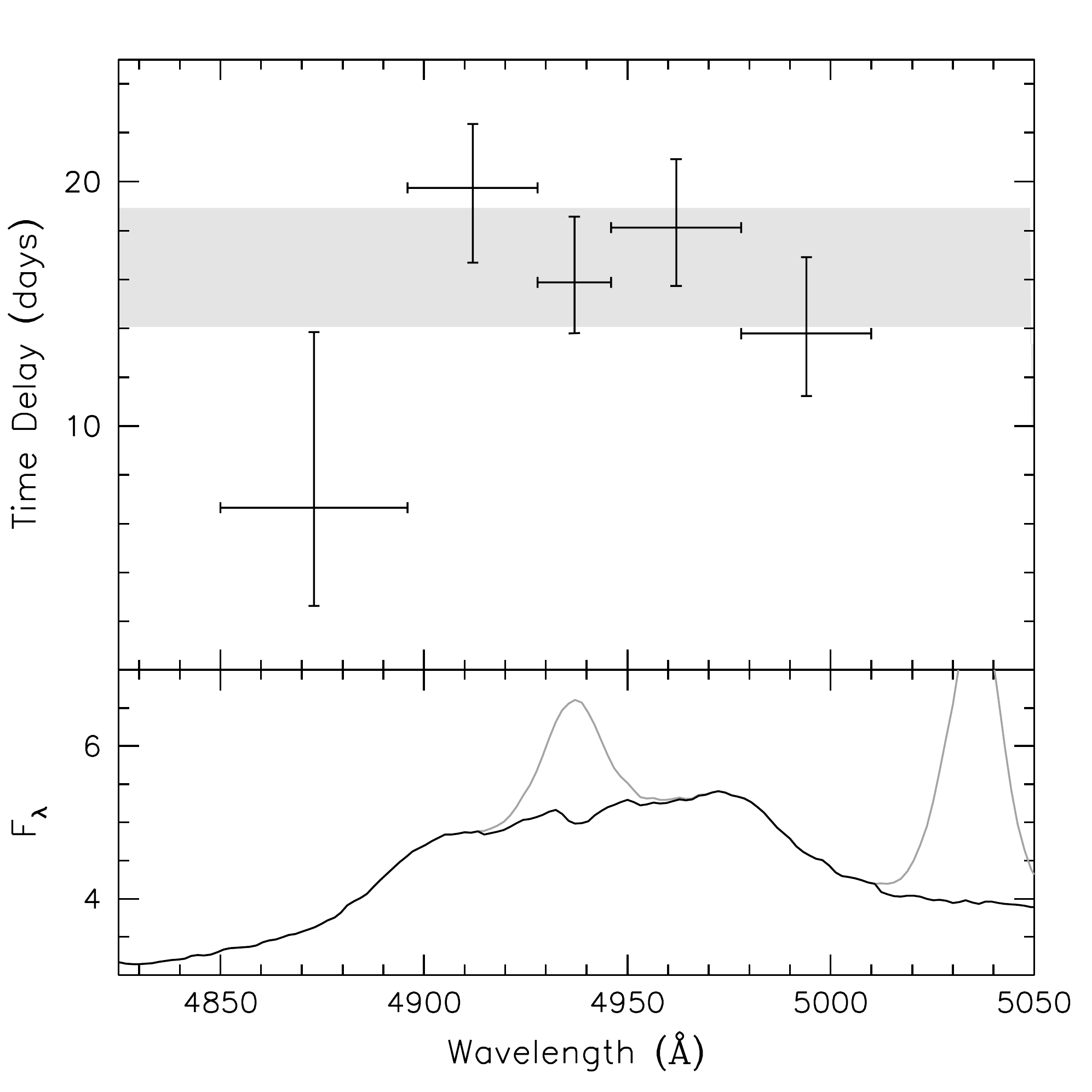}
    \caption{Velocity-resolved time delays (top) across the profile of the broad H$\beta$ emission line (bottom).  The narrow-line contributions to the mean spectrum are separately plotted in gray.}
    \label{fig:velres}
\end{figure}

\subsection{Velocity-Resolved Time Delays}

While a population-average scale factor $\langle f \rangle$ is generally adopted for reverberation masses, it was understood from the initial development of reverberation mapping that the response as a function of velocity across an emission line would provide information on the overall geometry and kinematics of the BLR gas \citep{blandford82}.  Mapping the BLR via reverberation would then allow the black hole mass to be directly constrained, without reliance on a scale factor. Initial attempts to constrain velocity-resolved time delays were generally hampered by observational uncertainties and systematics, but more recently there have been a growing number of successes (e.g., \citealt{bentz08,denney09c,grier12c,fausnaugh17,brotherton20,oknyansky21,bao22}). 

Therefore, in addition to measuring the time delay for the integrated H$\beta$ emission, we also investigated the possibility of constraining velocity-resolved time delays across the broad emission line profile.  Following the methods described in Section~\ref{sec:analysis}, we divided the H$\beta$ profile into {\bf five} velocity bins.  We created a light curve for each bin and measured the time delay with respect to the continuum light curve.

In Figure~\ref{fig:velres} we show the time delays ($\tau_{\rm cent}$)
measured as a function of velocity across the H$\beta$ emission line.
For comparison, the gray horizontal band marks the time delay and its
uncertainties for the integrated emission line. A slight M-shaped
pattern  may be apparent in the time delays, reminiscent of that
seen in the velocity-resolved time delays of the broad UV and optical
emission lines in NGC\,5548 \citep{derosa15,pei17}. Detailed analysis
of the velocity-resolved signatures in NGC\,5548 by \citet{horne21}
revealed the M-shaped pattern to be the signature of an inclined
Keplerian disk.  IC\,4329A has long been known
\citep{disney73,wilson79} as  one of the few nearby AGNs to show
clear double-peaked broad emission line profiles.
\citet{storchibergmann17} suggest that double peaks in optical broad
emission lines are more common among local Seyferts than generally
recognized, and they include NGC\,5548 among their sample.
Double-peaked broad emission lines are often interpreted as indicating
a disk-like broad line region structure, so the  potential
similarities between the velocity-resolved responses of NGC\,5548 and
IC\,4329A are perhaps unsurprising if they are both double-peaked
AGNs.  Nevertheless,  while velocity-resolved time delays may
  provide hints, a more thorough analysis is  required to
constrain the details of the BLR geometry in IC\,4329A.

\subsection{{\tt CARAMEL} Modeling}

There are two independent approaches for analyzing velocity-resolved reverberation responses: as an ill-posed inverse problem or through forward modeling.  The inverse approach attempts to extract the transfer function, which describes the time delay distribution as a function of velocity, directly from the data \citep{horne94,skielboe15,anderson21}.  Forward modeling, on the other hand, uses a framework of self-consistent models to explore the relevant parameter space and determine the set of models that best agree with the observations \citep{pancoast11,pancoast14a}.  Forward modeling has the advantage that the results are relatively simple to interpret, although it is limited by the flexibility and completeness of the models.  The inverse approach, on the other hand, makes few initial assumptions and is generally more flexible, but produces results that may be difficult to interpret and generally rely on comparison with models.

We explored the possibility of constraining more details regarding the geometry and kinematics of the BLR in IC\,4329A with the forward modeling code {\tt CARAMEL}, following a similar approach to that we previously employed \citep{bentz21c,bentz22a}.  In this case, most of the model parameters were not well constrained.  Parameters with modest constraints included the black hole mass, $\log ($\mbh/\msun$) =  7.64^{+0.53}_{-0.25}$ or $M_{\rm BH}=4^{+10}_{-2}\times10^7$\,\msun, and the average time delay,  $\tau_{\rm mean}=13.6^{+9.2}_{-3.0}$\,days, which agree with our results  in Sections~\ref{sec:delays} and \ref{sec:mbh}.  

\begin{figure}
    \centering
    \epsscale{1.15}
    \plotone{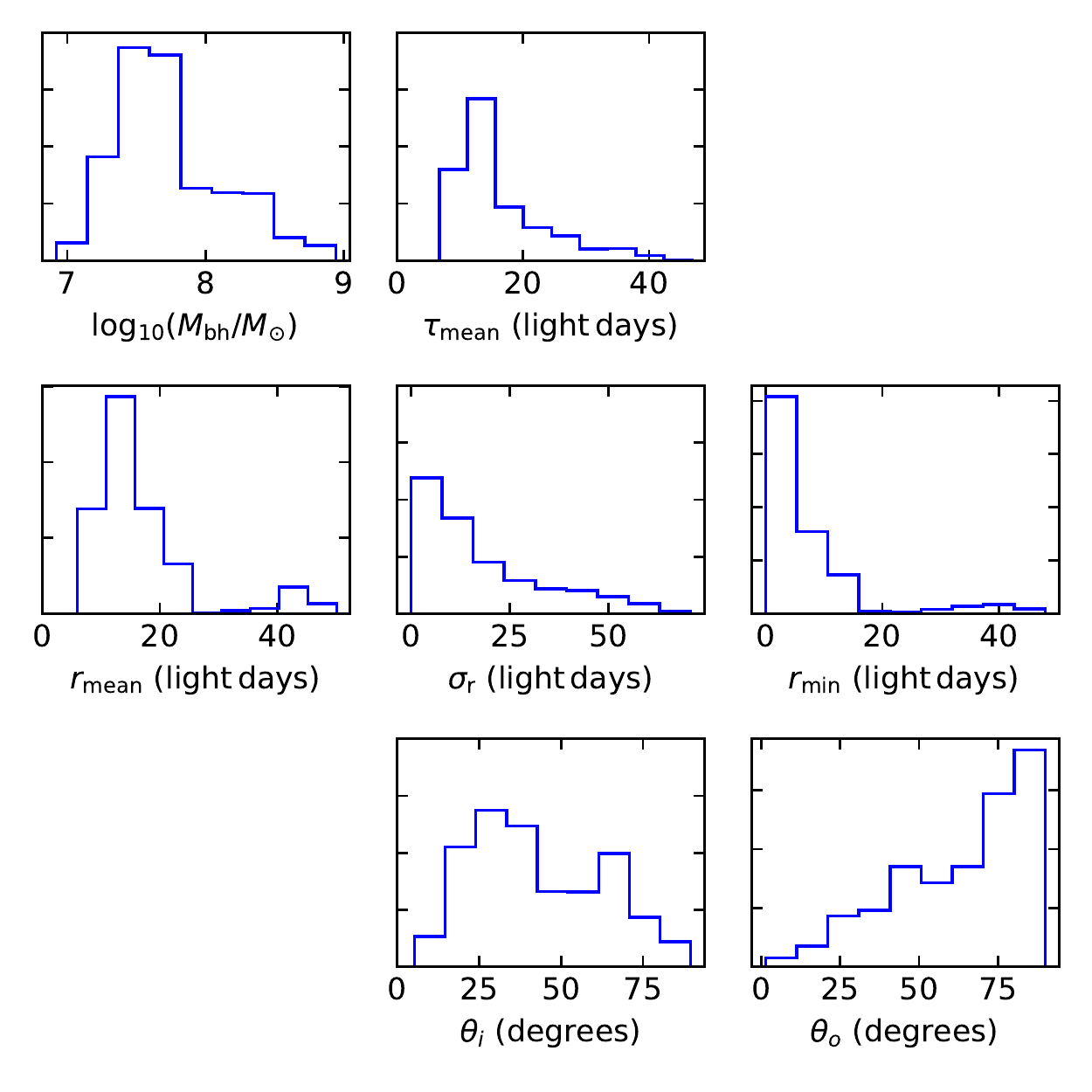}
    \caption{Posterior distributions of constraints on the H$\beta$-emitting BLR geometry.}
    \label{fig:posterior}
\end{figure}

While the dynamics of the system were not constrained, there were weak constraints on some of the geometric parameters of the H$\beta$-emitting BLR (Figure~\ref{fig:posterior}), suggesting an average radius of $r_{\rm mean}=14.2^{+7.2}_{-3.7}$\,light-days, with a minimum radius of $r_{\rm min}=3.1^{+8.5}_{-2.5}$\,light-days and a radial extent of $\sigma_{\rm r}=12.0^{+22.5}_{-8.0}$\,light-days.

Finally, modest constraints were derived for the inclination angle of
the BLR relative to our line of sight, $\theta_i = 40.1
^{+26.6}_{-18.1}$\,deg, as well as the opening angle of the BLR,
$\theta_o=69.2^{+15.3}_{-29.6}$\,deg, where $\theta_o=0\degr$ is a
thin disk and $\theta_o=90\degr$ is a sphere.  The inclination angle
suggests that the AGN system and the galaxy disk are indeed misaligned
by a significant amount ($\sim25\degr-70\degr$). Furthermore, we note
that the moderate inclination angle constrained by the models suggests
that the value of $\langle f \rangle$ adopted in Section~\ref{sec:mbh}
is unlikely to be an underestimate for this particular AGN
(cf.\ \citealt{williams18}).

Visualization of a random sampling of allowed BLR models included wide and puffy inclined annuli, filled inclined biconical structures, and inclined ``hamburger buns'' that resemble the intersection of a bicone with a shell.  Thus, while these models demonstrate consistency with several previous results including those that we present above, they are limited in their ability to elucidate a more clear picture of the BLR in IC\,4329A.  Additional work will be required to determine whether these limitations are a result of the quality of the reverberation data, which are a bit noisy but do show a clear velocity-resolved response, or are instead due to a mismatch between the modeling assumptions in {\tt CARAMEL} and the physical conditions of the BLR in IC\,4329A.

\section{Summary}

We have presented a new reverberation mapping campaign focused on the
AGN in the center of the edge-on spiral galaxy IC\,4329A.  With
photometric and spectroscopic monitoring covering 2022 Feb--July, we
were able to constrain the time delay of the broad H$\beta$ and
H$\gamma$ emission lines: $16.3^{+2.6}_{-2.3}$\,days for H$\beta$ and
$16.0^{+4.8}_{-2.6}$\,days for H$\gamma$.  \ion{He}{2} $\lambda 4686$
also exhibited variability during these observations, but the time
delay of $-0.6^{+3.9}_{-3.9}$\,days is unresolved and consistent with
zero days.  The H$\beta$ time delay is consistent with that expected
from the relationship between AGN luminosity and BLR radius for local
Seyferts, after correcting the AGN luminosity for the expected
$\sim2.4$\,mag of intrinsic extinction at 5100\,\AA.

Combining the time delay and width of H$\beta$, we find a black hole
mass of $M_{\rm BH}=6.8^{+1.2}_{-1.1}\times10^7$\msun\ when adopting
$\langle f \rangle = 4.8$.  Based on the detection of
velocity-resolved time delays across the H$\beta$ broad line profile,
we carried out forward modeling of the reverberation data presented
here.  Many of the model parameters were relatively unconstrained, but
the black hole mass and average H$\beta$ time delay agreed with our
measurements.  Finally, the models suggest that the BLR is indeed
severely misaligned with the disk of the host galaxy, in agreement
with expectations from the unified model for AGNs given that broad
emission lines are clearly visible in the spectrum of the nucleus of
IC\,4329A, and yet the galaxy is oriented edge-on to our line of
sight.

\begin{acknowledgements}
We thank the referees for suggestions that improved the presentation
of this work. We thank Tommaso Treu and Peter Williams for providing
the {\tt CARAMEL} modeling code as well as assistance in working with
it. MCB gratefully acknowledges support from the NSF through grant
AST-2009230. CAO was supported by the Australian Research Council
(ARC) through Discovery Project DP190100252. MV was supported by the
NSF through AST-2009122.

\end{acknowledgements}


\facility{LCOGT}

\software{LCO Reduction Pipeline \citep{mccully18}, JAVELIN \citep{zu11}, Galfit \citep{peng02,peng10}, CARAMEL \citep{pancoast14a} }

\bibliography{mbentz}{}
\bibliographystyle{aasjournal}

\end{document}